\documentclass[12pt]{article}
\usepackage[a4paper, margin=1in]{geometry}
\usepackage{graphicx}
\usepackage{amsmath}
\usepackage{cite}
\usepackage{hyperref}
\usepackage{titlesec}

\titleformat{\section}{\normalfont\Large\bfseries}{\thesection.}{1em}{}
\titleformat{\subsection}{\normalfont\large\bfseries}{\thesubsection.}{1em}{}

\title{Integrated Experimental and Numerical Investigations on the Thermo--Hydro--Mechanical Behavior of Clays and Argillaceous Rocks: A Perspective}

\author{Saeed Tourchi\thanks{Corresponding author: \href{mailto:saeed.tourchi@uni.lu}{saeed.tourchi@uni.lu}}\\
\small Computational Soil Mechanics and Foundation Engineering (COMPSOIL),\\
\small Department of Engineering, University of Luxembourg, Luxembourg}

\date{}

\begin{document}
\maketitle

\begin{abstract}
Over the past decade a coordinated research programme has been carried out to
characterise, model and predict the coupled thermo–hydro–mechanical behaviour
of clays and argillaceous rocks across laboratory, intermediate and field scales.
The work spans high-precision triaxial tests on compacted bentonites and stiff
claystones, multi-year heating experiments in underground research laboratories,
and full three-dimensional finite-element simulations of repository cells, energy
piles and climate-sensitive slopes.  A family of critical-state based constitutive
models is presented that unifies temperature-dependent elasto-plasticity,
viscoplastic creep, suction hysteresis and structure degradation within a single
thermodynamically consistent framework.  The models are implemented in an
implicit, fully coupled THM code and calibrated with a Bayesian workflow using
sensor data from instrumented tunnels, boreholes and surface observatories.

Key outcomes include: (i) explanation of anisotropic tunnel convergence driven
by delayed thermo-swelling; (ii) quantification of permeability changes in
excavation-damaged zones subjected to sustained heating; (iii) identification of
critical suction thresholds for desiccation cracking and heterogeneous rebound in
expansive strata; and (iv) demonstration of cyclic thermal effects on the shaft
capacity of energy foundations.  The resulting design charts and digital twins
provide practical guidance for nuclear-waste repository layouts, geothermal
retaining walls and slope-stability assessments under a warming climate.

The paper synthesises these findings into a transferable modelling strategy that
couples physics-based formulations with data-driven updating, setting the stage
for next-generation THM analyses that incorporate chemical reactions, real-time
monitoring and probabilistic performance evaluation.\\


\end{abstract}

\section{Introduction}

Understanding the thermo–hydro–mechanical (THM) behaviour of clays and claystones is critical for the long-term performance of nuclear repositories, energy geostructures and natural slopes.  Over the past decade our research group has developed a suite of constitutive models, laboratory protocols and field-scale simulations that progressively captured the coupled influence of temperature, suction and stress on these low-permeability geomaterials.

The effort began with critical-state formulations for unsaturated expansive soils \cite{tourchi2015unsat,hamidi2015saturated}, soon extended to incorporate suction‐dependent hardening and temperature effects in bonded and structured clays \cite{hamidi2016unsat,hamidi2017structured}.  A comprehensive doctoral study formalised these advances into a unified THM framework for argillaceous rocks \cite{tourchi2020thesis}.  Subsequent work introduced temperature-dependent creep and anisotropic permeability, yielding a general thermomechanical model for hard clays and weak rocks \cite{tourchi2020e3s} and, more recently, a viscoplastic slip-surface model for cyclic heating problems \cite{tourchi2025geotech}.  An extended hard-soils–weak-rocks formulation has also been benchmarked against deep-excavation data \cite{tourchi2025geotechnique}. Model credibility has been established through a programme of laboratory and intermediate-scale experiments.  These include wetting–drying tests on compacted bentonite and Iranian expansive soil \cite{ghandilou2023unsat}, cyclic hydration of Boom clay using the Barcelona Expansive Model \cite{ghandilou2023boom}, and thermo-mechanical tests on silty sandy clays to verify temperature-related stiffness degradation \cite{hoseinimighani2023numge, mighani2025impact, hoseinimighani2026thermal}. Complementary element testing addressed the temperature dependence of the residual shear strength of bentonite \cite{tourchi2023bentonite} and the thermal failure of sand--clay mixtures \cite{tourchi2024thermalfailure}.

Parallel field campaigns have provided essential validation data.  Coupled analyses of a full-scale steel-lined and an unsupported micro-tunnel in Callovo-Oxfordian (COx) claystone demonstrated the need to consider anisotropic swelling and viscoplastic convergence \cite{tourchi2019coupled,tourchi2019thermo}.  A three-year in-situ heating experiment in the COx formation further confirmed the model’s ability to reproduce temperature rise, pore-pressure transients and delayed expansion \cite{tourchi2021insitu}, while a companion study examined cell feasibility and host-rock response for high-level-waste disposal \cite{bumbieler2021hlw}.  The framework was later applied to deep-excavation problems in claystone using large-strain numerical methods \cite{tourchi2023numge} and to non-isothermal excavation-damaged zones around HLW galleries \cite{tourchi2024clayconf}, and to a TBM tunnel driven in Opalinus Clay, where thermal coupling governs the long-term lining pressure and the growth of the excavation damaged zone \cite{tourchi2026opalinus}. 
Surface-exposed clays present additional challenges stemming from soil–atmosphere interactions.  Recent simulations captured cracking and moisture dynamics in desiccated strata \cite{jabbarzadeh2024gete,jabbarzadeh2024soil,sadeghi2024enggeo} and reproduced the THM evolution of a weathered black-shale slope at the Požáry test site \cite{tourchi2024pozary}, building on earlier work relating temperature to slope stability in temperate climates \cite{scaringi2022temperature}.  Complementary studies on geothermal foundations highlight coupled heating effects on deep energy systems \cite{tourchi2025geotech,tourchi2025piles,tourchi2026shaft}.

In the sections that follow, Section 2 details the constitutive framework and its calibration; Section 3 reviews in-situ heating and excavation studies; Section 4 examines soil–atmosphere coupling and slope processes; and Section 5 outlines future research directions for energy sheet-pile walls, tunnels and climate-resilient infrastructure.

\section{Constitutive Modeling for Saturated and Unsaturated Clays}

Over the past decade, a series of contributions have been made to the development of thermo-hydro-mechanical (THM) constitutive models aimed at describing the complex behavior of saturated and unsaturated clays. The work originated from early investigations into the thermal response of unsaturated expansive soils and gradually expanded to address temperature-dependent plasticity, structure degradation, viscoplasticity, and anisotropy in stiff clays and claystones. The first step involved modifying the Barcelona Basic Model to include non-isothermal effects, leading to a generalized THM model for unsaturated expansive soils that could simulate irreversible volume changes during heating and drying–wetting cycles \cite{tourchi2015unsat, hamidi2016unsat}. This model accounted for thermal expansion and contraction, thermal hardening, and the collapse of soil structure under suction loss. Experimental results on compacted bentonite and other natural soils were used to calibrate and validate the formulation.

Subsequent developments included the introduction of structure and bonding effects into the constitutive framework. In \cite{hamidi2017structured}, a model was proposed to capture the behavior of structured clays under thermal and hydraulic perturbations, including degradation of bonding with mechanical loading and temperature increase. This extended the model’s capability to simulate stress-path dependent softening and hardening behavior. The culmination of this line of research was presented in the doctoral thesis \cite{tourchi2020thesis}, where the non-isothermal elasto-viscoplastic behavior of stiff claystones was formalized into a unified THM framework. The model integrated temperature-dependent creep, structure degradation, anisotropic strength, and permeability evolution, calibrated using data from both laboratory heating tests and in situ observations.

The THM model was then implemented into finite element software and applied to simulate large-scale problems such as the micro-tunnel excavated in Callovo-Oxfordian (COx) claystone \cite{tourchi2019coupled, tourchi2019thermo}. These simulations revealed the critical role of thermal loads on tunnel convergence and pore pressure dissipation. Later works focused on improving the coupling between heat flow, hydraulic flow, and stress evolution by introducing rate-dependent formulations, as validated by comparisons to field data from radioactive waste repository zones \cite{tourchi2021insitu}. The model was further applied to study the response of energy geostructures and the behavior of desiccated surface soils. For instance, in \cite{tourchi2020e3s}, the constitutive framework was adapted to evaluate the degradation of shaft resistance in energy piles due to cyclic thermal loading, while in \cite{jabbarzadeh2024gete, ghandilou2023unsat, ghandilou2023boom, tourchi2025piles} the focus shifted to cracking and volumetric instability during drying–wetting cycles.

Recent contributions have concentrated on extending the THM framework to analyze energy tunnel applications and energy sheet pile walls \cite{tourchi2023numge, tourchi2025geotech, tourchi2025holistic, zamani2023pumped}, emphasizing the need to accurately model the interface behavior and the progressive degradation of shear strength under coupled loading. The evolution of this constitutive modeling work reflects a gradual refinement from phenomenological models for unsaturated soils to a fully coupled, multiphysics framework tailored for low-permeability geomaterials subjected to thermal loading and long-term deformation. This modeling architecture provides a consistent basis for interpreting lab experiments, simulating field-scale responses, and designing geo-energy systems and nuclear repositories.

\section{In Situ Heating Tests for Deep Geological Repositories}

Field-scale heating experiments are essential to evaluate the long-term behavior of engineered barriers and surrounding claystone formations under repository-relevant conditions. In particular, the Callovo-Oxfordian (COx) claystone has been the focus of several in situ heating campaigns at the Meuse/Haute-Marne Underground Research Laboratory (URL) in France. These experiments provide unique datasets for validating numerical models and understanding the multi-physics interactions controlling the performance of high-level radioactive waste (HLW) disposal systems.

Our numerical efforts have targeted both thermal response and deformation characteristics observed during these experiments. In \cite{tourchi2021insitu,bumbieler2021hlw}, we used a fully coupled THM framework to simulate the long-duration heating phase and its effect on pore pressure buildup, thermal expansion, and stress redistribution in COx. The model incorporated anisotropic thermal conductivity, temperature-dependent permeability, and viscoplastic swelling behavior. Simulations successfully reproduced key trends in thermal profiles, pore pressure evolution, and radial displacements measured by sensors embedded in the host rock. This provided strong evidence for the role of time-dependent swelling and anisotropic flow during thermal loading. The in situ heating analyses were further extended in \cite{tourchi2024clayconf}, where the behavior of the excavation damaged zone (EDZ) was studied under thermal gradients representative of HLW emplacement. These simulations showed that EDZ permeability and saturation can evolve significantly under long-term heating, affecting the sealing properties of the disturbed zone. The coupled model captured transitions between partially saturated and saturated states in the EDZ and illustrated the gradual development of thermal stresses leading to damage reactivation.

Earlier works \cite{tourchi2019coupled, tourchi2019thermo} focused on the long-term convergence of micro-tunnels in COx, both steel-lined and unsupported. The THM model simulated anisotropic deformation over several years and highlighted the delayed buildup of radial pressures due to combined swelling and temperature effects. Comparisons with in situ measurements validated the model’s predictive capabilities and emphasized the importance of considering time-dependent properties and thermo-elastic-plastic coupling in tunnel design. These in situ applications were developed and refined over the course of the author’s doctoral work \cite{tourchi2020thesis}, which systematically integrated field data, constitutive model calibration, and advanced finite element implementation. The model provided a consistent framework to simulate both repository-scale experiments and component-scale laboratory tests, reinforcing its applicability for performance assessment studies.

Our work across multiple test campaigns and simulation scenarios confirms that reliable performance predictions for HLW disposal systems require explicit coupling of thermal, hydraulic, and mechanical processes. These findings support the development of advanced THM design tools for underground repositories in claystone formations.

\section{Soil–Atmosphere Interactions: Desiccation Cracking and Slope Processes}

Near-surface soils are exposed to climate-driven variations in temperature, humidity, and rainfall, which induce complex thermo-hydro-mechanical (THM) cycles. These interactions between soil and atmosphere can significantly alter the mechanical behavior of expansive clays, leading to fissuring, strength loss, and long-term instability in infrastructure and slopes. We have applied coupled THM numerical models to simulate the desiccation of unsaturated clays under drying–wetting cycles. In \cite{jabbarzadeh2024gete}, we investigated the evolution of suction, stiffness degradation, and crack initiation during desiccation. The results were validated against laboratory-scale tests and revealed the existence of critical suction thresholds beyond which cracking becomes irreversible. These simulations demonstrated that even moderate atmospheric drying can generate localized tensile zones that lead to fissure propagation, particularly in low-permeability clays. Complementary studies modeled the volumetric behavior of expansive soils under cyclic hydric conditions using the Barcelona Expansive Model (BExM) \cite{ghandilou2023unsat}. This framework was extended in \cite{ghandilou2023boom} to simulate the wetting and drying behavior of Boom clay. Numerical analyses reproduced the accumulation of irreversible strains and shifts in pore structure due to repeated cycles. The simulations highlighted the role of microstructural collapse and suction hysteresis in long-term damage accumulation, which in turn influences surface heave and settlement behavior.

In natural slope environments, additional complexity arises from the interaction of thermal and oxidative processes with moisture changes. Our investigation at the Požáry test site integrated in situ meteorological and soil data with numerical simulations to assess slope performance under seasonal and long-term environmental loading \cite{tourchi2024pozary}. The results illustrated that shallow failures can be triggered by thermal drying and oxidation-induced weakening of clayey layers. The THM model captured both short-term pore pressure responses and long-term strength degradation mechanisms, which are not explained by purely hydromechanical models. 
These contributions underscore the necessity of incorporating soil–atmosphere interactions into geotechnical hazard assessment and infrastructure design. They also highlight the critical role of climate-resilient modeling frameworks capable of capturing the complex feedback between environmental drivers and geomaterial behavior.

\section{Outlook and Future Directions}

The research trajectory presented here opens several strategic avenues for advancing thermo–hydro–mechanical (THM) modelling and its field deployment.  One priority is the translation of laboratory-validated constitutive laws into fully coupled three-dimensional analyses for energy tunnels, geothermal retaining structures and other heat-exchanging geosystems \cite{tourchi2025geotech,tourchi2023numge}.  These applications demand rapid thermal cycling, forced ventilation and transient groundwater flow—conditions markedly different from the slower gradients observed in nuclear repositories.  Extending the current frameworks to capture such high-frequency loads will require robust time-integration schemes, refined interface elements and efficient parallel solvers.  Planned collaborations with infrastructure operators should supply continuous monitoring data for real-time model updating and decision support.

A second line of inquiry concerns chemo-thermo-hydro-mechanical coupling, particularly the role of oxidation, carbonation and mineral dissolution in long-term barrier performance \cite{tourchi2024clayconf}.  Preliminary laboratory tests indicate that oxidative micro-cracking may accelerate hydraulic conductivity increases in clay-rich rocks; incorporating such effects into the viscoplastic framework could explain several in-situ anomalies noted during extended heating phases \cite{tourchi2021insitu}.  Coupling geochemical reaction networks with the existing THM code base, while preserving numerical stability, will be addressed through operator-splitting strategies and adaptive time stepping.  These enhancements should bolster confidence in life-cycle assessments for both radioactive-waste cells and geo-energy boreholes.

Ongoing work also targets upscaling and regional integration.  Repository-scale models of Callovo-Oxfordian (COx) claystone now span hundreds of metres and incorporate rock anisotropy, swelling convergence and excavation-damaged-zone evolution \cite{tourchi2019coupled,bumbieler2021hlw}.  The next step is to embed those high-resolution domains within basin-scale hydrogeological simulators to evaluate far-field perturbations—thermal plumes, pore-pressure transients and gas migration—over millennial timescales.  Multiscale coupling will leverage domain-decomposition techniques and reduced-order surrogates calibrated on detailed FE results, accelerating probabilistic safety-case evaluations.

Sensor-rich field observatories remain essential for closing knowledge gaps.  Planned deployments include fibre-optic distributed temperature sensing in new deep-borehole heat-exchangers, wireless pressure arrays along energy sheet-pile walls, and photogrammetric crack-mapping on instrumented slopes \cite{tourchi2024pozary}.  These datasets will feed machine-learning pipelines for parameter inversion and anomaly detection, complementing physics-based models with data-driven insights.  Hybrid digital twins—combining reduced THM solvers and neural surrogates—are envisioned as real-time tools for infrastructure owners.

Finally, the group will continue refining constitutive laws to encompass rate-dependent damage, anisotropic creep and partially saturated fracture flow, as recently trialled in the hard-soil–weak-rock formulation for deep excavations \cite{tourchi2025geotechnique}.  Emphasis will be placed on thermodynamic consistency, efficient sensitivity analysis and open-source dissemination to maximise community uptake.  Collectively, these initiatives aim to deliver predictive, adaptable and field-verified THM models that underpin resilient design across underground storage, energy geostructures and climate-sensitive slopes.

\bibliographystyle{IEEEtran}
\bibliography{thm_papers}

\end{document}